# Advance of Mercury Perihelion Explained by Cogravity

C. J. de Matos[*]

*Coimbra University, 3000 Coimbra, Portugal*

M. Tajmar[†]

*Vienna University of Technology, 1040 Vienna, Austria*

**Abstract**

The theory of General Relativity explaines the advance of Mercury perihelion using space curvature and the Schwartzschild metric. We demonstrate that this phenomena can also be interpreted due to the cogravitational field produced by the apparent motion of the Sun around Mercury giving exactly the same estimate as derived from the Schwartzschild metric in general relativity theory. This is a surprising result because the estimate from both theoretical approaches match exactly the measured value. The discussion and implications of this result is out of the scope of the present work.

---
[*] Graduate Associate, Present Address: European Space Research and Technology Centre ESA-ESTEC, P.O. Box 299, 2200 AG Noordwijk, The Netherlands. E-mail: cdematos@estec.esa.nl
[†] Postdoctoral Associate, Present Address: European Space Research and Technology Centre ESA-ESTEC, P.O. Box 299, 2200 AG Noordwijk, The Netherlands. E-mail: tajmar@bigfoot.com

## The Theory of the Gravitational and Cogravitational Fields

Doing the following substitution in Maxwell's electromagnetic (EM) field theory we obtain the theory of the gravitational and cogravitational fields also designated as the theory of gravito-cogravitism (GC). For a detailed analysis, the reader is referred to the literature (e.g. Jefimenko, 1992). All important expressions are summarized in Table 1. The cogravitational field $\vec{K}$ is for the gravitational field $\vec{g}$ what the magnetic field $\vec{B}$ is for the electric field $\vec{E}$.

| Electromagnetism | Gravitational & Cogravitational Fields |
|---|---|
| $q$ (electric charge) | $m$ (mass) |
| $\rho$ (volume charge density) | $\rho_m$ (volume mass density) |
| $\sigma$ (surface charge density) | $\sigma_m$ (surface mass density) |
| $\lambda$ (line charge density) | $\lambda_m$ (line mass density) |
| $\vec{j}$ (electric current density) | $\vec{j}_m$ (mass current density) |
| $\vec{E}$ | $\vec{g}$ |
| $\vec{B}$ | $\vec{K}$ |
| $\varepsilon_0$ | $-1/4\pi G$ |
| $\mu_0$ | $-4\pi G/c^2$ |
| $-1/4\pi\varepsilon_0$ or $-\mu_0 c^2/4\pi$ | $G$ (the universal gravitational constant) |

**Table 1** Corresponding Electromagnetism and Gravitocogravitism Symbols and Constants

The following important results summarise the theory of Gravitocogravitism:

**a)** The local equations of the Gravitocogravitic field in vacuum are

$$\nabla \vec{g} = -4\pi G \rho_m \tag{1}$$

$$\nabla \vec{K} = 0 \tag{2}$$

$$\nabla \times \vec{g} = -\frac{\partial \vec{K}}{\partial t} \tag{3}$$

$$\nabla \times \vec{K} = -\frac{4\pi G}{c^2} \vec{j}_m + \frac{1}{c^2} \frac{\partial \vec{g}}{\partial t}. \tag{4}$$

where $\vec{j}_m = \rho_m \vec{v}$ is the mass current density and $\rho_m$ is the density of mass.

**b)** The law that gives us the value of the cogravitational field $\vec{K}$ are a point $\vec{r}$ created by the motion of a point mass *m*, is

$$\vec{K} = -\frac{G}{c^2}\frac{m\vec{v} \times \vec{r}}{r^3} \tag{5}$$

and the associated cogravitational vector potential is

$$\vec{A} = -\frac{G}{c^2}\frac{m\vec{v}}{r} \tag{6}$$

**c)** The total force acting on a particle of mass *m* in a gravitational and cogravitational field is:

$$\vec{F} = m(\vec{g} + \vec{v} \times \vec{K}) \tag{7}$$

**d)** The gravitational and the cogravitational forces acting upon two masses $m_1$ and $m_2$ moving parallel to each other with the same velocity $\vec{v}$ (with respect to a reference frame linked to the laboratory) are respectively:

$$F_{grav} = G\frac{m_1 m_2}{r^2} \tag{8}$$

$$F_{cograv} = \frac{G}{c^2}\frac{m_1 m_2}{r^2}v^2. \tag{9}$$

The relative intensity of these two forces can be evaluated to be:

$$\frac{F_{cograv}}{F_{grav}} = \left(\frac{v}{c}\right)^2. \tag{10}$$

From this last result, we see clearly that the cogravitational force is much weaker than the gravitational force when the masses are moving with velocities much lower than the speed of light. In a common Earth laboratory experiment the velocities involved are much lower than *c* and the gravitational forces created by the masses used in the experiments are hardly detectable due to their very weak value. These two facts considered simultaneously explain why the cogravitational field has never been detected so far in an Earth laboratory experiment (e.g. Braginski et al, 1977).

**e)** In a GC wave we have the following relation between the gravitational and the cogravitational field:

$$K = \frac{g}{c} \tag{11}$$

This last result shows that in a GC wave the gravitocogravitational field is $c^{-1}$ times weaker than the gravitational field. This is the reason why such waves are so difficult to be detected.

**f)** The relativistic motion of a particle of proper mass $m_0$ in a cogravitational field can be extracted from the following Lagrangian (Rocard, 1992)

$$L = \left(1 - \frac{v^2}{c^2}\right)^{-1/2} \left[m_0 c^2 - m_0 f + m_0 \vec{v} \vec{A}\right] \tag{12}$$

where $f$ is the Newtonian gravitational potential. Using the Lagrangian in Equ. (12) we will calculate the advance of the Mercury perihelion.

It is well known that for weak gravitational fields, the linearized form of General Relativity turns out to be very similar to the theory of GC (Forward, 1961). However, during the linearization process, one has an arbitrary choice regarding the value of the speed of propagation of the GC field and the way we express the gravitational Lorentz force law as well as the cogravitational potential energy. Indeed, either of the two following possibilities are acceptable:

| Speed of propagation of the GC field | Gravitational Lorentz force law | Cogravitational potential energy |
|---|---|---|
| $c$ | $\vec{F}_k = 4m\vec{v} \times \vec{K}$ | $E = -4m_0\vec{v}\vec{A}$ |
| $c/2$ | $\vec{F}_k = m\vec{v} \times \vec{K}$ | $E = -m_0\vec{v}\vec{A}$ |

This shows clearly that the linearized theory of GR is not perfectly isomorphic (Peng, 1983, Tajmar et al, 2000) with electromagnetism which is commonly understood as a limitation of linearized GR.

**The Advance of Mercury Perihelion**

Let us consider the general motion of a body of mass $m_0$ moving around another body of mass $M_0$ such that $m_0 << M_0$. When we write the relativistic Lagrangian of this system we have to take into account the cogravitational potential energy created by the apparent motion of $M_0$ with respect to $m_0$. In our Equ. (12), we can express the Newtonian gravitational potential created by the mass $M_0$ as

$$f = -G\frac{M_0}{r} \tag{13}$$

Moreover, $\vec{v}$ is the velocity of $m_0$ with respect to $M_0$ and $\vec{A}$ is the cogravitational vector potential created by the apparent motion of $M_0$ with respect to $m_0$ ($-\vec{v}$). It is given by

$$\vec{A} = -\frac{G}{c^2}\frac{M_0}{\sqrt{1-\left(\frac{v}{c}\right)^2}}\frac{(-\vec{v})}{r} \tag{14}$$

By substitution of Equ. (13) and (14) in Equ. (12) we have:

$$L = \frac{m_0 c^2}{\sqrt{1-\left(\frac{v}{c}\right)^2}} + \frac{GM_0 m_0}{r\sqrt{1-\left(\frac{v}{c}\right)^2}} + \frac{GM_0 m_0}{c^2\sqrt{1-\left(\frac{v}{c}\right)^2}} \frac{v^2}{r} \quad (15)$$

If we consider $v \ll c$ the Lagrangian in Equ. (15) is transformed into

$$L = m_0 c^2 + \frac{1}{2} m_0 v^2 + \frac{GM_0 m_0}{r} + \frac{3}{2}\frac{GM_0 m_0}{c^2}\frac{v^2}{r} \quad (16)$$

where the term 3/2 comes from the summation of the taylor develoment of the gravitic potential energy with the cogravitational potential energy. We see that the velocity contained in the cogravitational potential energy term can be the velocity of $m_0$ with respect to $M_0$, i.e., $\vec{v}$ or the velocity of $M_0$ with respect to $m_0$, i.e., $-\vec{v}$, because we have a squared velocity. Therefore if we define

$$\vec{\tilde{v}} = -\vec{v} \quad (17)$$

We can then write Equ. (16) as

$$L = m_0 c^2 + \frac{1}{2} m_0 v^2 + \frac{GM_0 m_0}{r} + \frac{3}{2}\frac{GM_0 m_0}{c^2}\frac{\tilde{v}^2}{r} \quad (18)$$

We see that the kinetic energy does not contain $\tilde{v}$ because we want to establish the equations of motion of $m_0$ with respect to the reference frame attached to $M_0$, but the cogravitational field felt by $m_0$ is due to the apparent motion of $M_0$ with respect to the reference frame attached to $m_0$ (see Figure 1).

Using polar coordinates we can write the velocity as

$$\vec{v} = \dot{r}\hat{e}_r + r\dot{q}\,\hat{e}_q \quad (19)$$

$$\vec{\tilde{v}} = \dot{\tilde{r}}\,\hat{\tilde{e}}_r + \tilde{r}\dot{\tilde{q}}\,\hat{\tilde{e}}_q \quad (20)$$

by substitution of Equ. (19) and (20) into (18)

$$L = m_0 c^2 + \frac{1}{2} m_0 \left( \dot{r}^2 + r^2 \dot{\boldsymbol{q}}^2 \right) + \frac{GM_0 m_0}{r} + \frac{3}{2} \frac{GM_0 m_0}{c^2} \frac{\left( \dot{\tilde{r}}^2 + \tilde{r}^2 \dot{\tilde{\boldsymbol{q}}}^2 \right)}{r} \tag{21}$$

Noting that in our case

$$\frac{d}{dt} \frac{\partial L}{\partial \dot{x}} = \frac{\partial L}{\partial x} \Leftrightarrow \frac{d}{dt} \frac{\partial L}{\partial \dot{\tilde{x}}} = \frac{\partial L}{\partial \tilde{x}} \tag{22}$$

where $x, \dot{x}$ are respectively, generalised coordinates and generalised velocities, we can write the Lagrange equations as:

$$\frac{d}{dt} \frac{\partial L}{\partial \dot{\boldsymbol{q}}} = \frac{\partial L}{\partial \boldsymbol{q}} \tag{23}$$

$$\frac{d}{dt} \frac{\partial L}{\partial \dot{r}} = \frac{\partial L}{\partial r} \tag{24}$$

From Equ. (23) we have the angular momentum $\ell$ that we consider as being approximately

$$\ell \approx m_0 r^2 \dot{\boldsymbol{q}} \tag{25}$$

which gives us

$$\dot{\boldsymbol{q}} = \dot{\tilde{\boldsymbol{q}}} = \frac{\ell}{r^2 m_0} \tag{26}$$

From Equ. (24) we have

$$m_0 \ddot{r} + \frac{3GM_0 m_0}{c^2} \frac{\ddot{\tilde{r}}}{r} - \frac{3GM_0 m_0}{c^2} \frac{\dot{r} \dot{\tilde{r}}}{r^2} = m_0 r \dot{\boldsymbol{q}}^2 - \frac{GM_0 m_0}{r^2} - \frac{3}{2} \frac{GM_0 m_0}{c^2} \frac{\dot{\tilde{r}}^2}{r^2} + \frac{3}{2} \frac{GM_0 m_0}{c^2} \dot{\tilde{\boldsymbol{q}}}^2 \tag{27}$$

and making the following substitution of the variable

$$u = \frac{1}{r} \qquad (28)$$

And

$$\tilde{u} = -\frac{1}{r} = \frac{1}{\tilde{r}} \qquad (29)$$

we finally obtain after an extensive calculation

$$\frac{d^2 u}{d\mathbf{q}^2} + u = \frac{GM_0 m_0}{\ell^2} - \frac{3}{2}\frac{GM_0}{c^2} u^2 - 3\frac{GM_0}{c^2} u \frac{d^2 \tilde{u}}{d\tilde{\mathbf{q}}^2} + \frac{3}{2}\frac{GM_0}{c^2}\left(\frac{d\tilde{u}}{d\tilde{\mathbf{q}}}\right)^2 - 3\frac{GM_0}{c^2}\frac{d\tilde{u}}{d\tilde{\mathbf{q}}}\frac{du}{d\mathbf{q}} \qquad (30)$$

From Equ. (30) we see that the equation of motion of a planet in the gravitocogravitic field theory differs from the Newtonian equation by the sum of quadratic terms:

$$R(u,\tilde{u}) = \left( -\frac{3}{2}\frac{GM_0}{c^2} u^2 - 3\frac{GM_0}{c^2} u \frac{d^2 \tilde{u}}{d\tilde{\mathbf{q}}^2} + \frac{3}{2}\frac{GM_0}{c^2}\left(\frac{d\tilde{u}}{d\tilde{\mathbf{q}}}\right)^2 - 3\frac{GM_0}{c^2}\frac{d\tilde{u}}{d\tilde{\mathbf{q}}}\frac{du}{d\mathbf{q}} \right) \qquad (31)$$

We can solve this equation by approximation, making the substitution of $u$ and $\tilde{u}$ inside the quadratic terms (31) by the solution $u_0$ (and the associated $\tilde{u}_0$) of the approximate equation:

$$\frac{d^2 u}{d\mathbf{q}^2} + u = \frac{GM_0 m_0}{\ell^2} \equiv \frac{1}{p} \qquad (32)$$

This means that we have two possible ways to solve Equ. (30):

1.)

$$u_0 = \frac{1 - e\cos(\mathbf{q} - \mathbf{q}_0)}{p} \qquad (33)$$

$$\tilde{u}_0 = \frac{1 + e\cos(\mathbf{q} - \mathbf{q}_0)}{p} \qquad (34)$$

or 2.)

$$u_0 = \frac{1 + e\cos(\mathbf{q} - \mathbf{q}_0)}{p} \qquad (35)$$

$$\tilde{u}_0 = \frac{1 - e\cos(\mathbf{q} - \mathbf{q}_0)}{p} \qquad (36)$$

We have a phase difference of $\mathbf{p}$ between $u_0$ and $\tilde{u}_0$ because we are considering simultaneously the relative motion of $M_0$ with respect to $m_0$ and *vice versa* (as shown in Figure 1).

Then, if we consider case (1.), by neglecting the terms that contain $e^2$ and noting that $\frac{3}{2}\frac{GM_0}{c^2 p} \ll 1$ we find that

$$\frac{d^2 u}{d\mathbf{q}^2} + u = \frac{1}{p} - 6\frac{GM_0}{c^2 p} e\cos(\mathbf{q} - \mathbf{q}_0) \qquad (37)$$

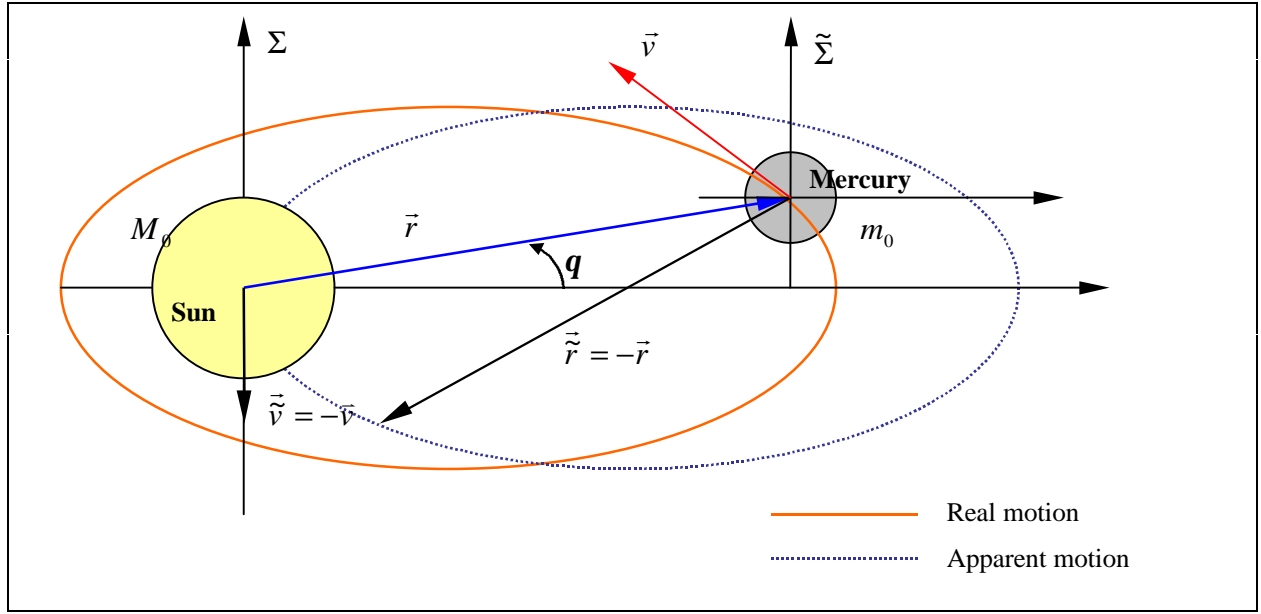

**Figure 1** The motion of Mercury around the Sun and the apparent motion of the Sun around Mercury

Introducing the new variable $U$

$$U = u - \frac{1}{p} \tag{38}$$

we obtain

$$\frac{d^2U}{d\boldsymbol{q}^2} + U \approx \frac{6GM_0}{c^2 p} U \tag{39}$$

From this we obtain

$$\frac{d^2U}{d\boldsymbol{q}^2} + \boldsymbol{a}^2 U \approx 0 \tag{40}$$

$$\boldsymbol{a}^2 \equiv 1 - \frac{6GM_0}{c^2 p} \tag{41}$$

The solution of Equ. (40) is

$$u = \frac{1}{r} = \frac{1 + e\cos[\boldsymbol{a}(\boldsymbol{q} + \boldsymbol{q}_0)]}{p} \tag{42}$$

as $e < 1$ and $\boldsymbol{a} \approx 1$. The trajectory will be very close to an ellipse with the major axis rotating in the direct sense at each revolution, by an angle of

$$\boldsymbol{d} = \frac{2\boldsymbol{p}}{\boldsymbol{a}} - 2\boldsymbol{p} = 2\boldsymbol{p}\left\{\left(1 - \frac{6GM_0}{c^2 p}\right)^{-1/2} - 1\right\} \approx \frac{6\boldsymbol{p}GM_0}{c^2 p} \tag{43}$$

As a function of the eccentricity $e$ and of the semi major axis $a$, we get

$$\boldsymbol{d} = \frac{6\boldsymbol{p}GM_0}{c^2 a(1 - e^2)} \tag{44}$$

which is exactly the result which we obtain from the theory of general relativity using the Schwartzschild metric. This effect is a maximum for the planet Mercury. In this particular case, $a = 0.579 \times 10^{11} m$, and $e = 0.206$ we find $\boldsymbol{d} = 0.104''$ per revolution.

If we consider Case 2 we obtain a negative advance of the perihelion,

$$\boldsymbol{d} = -\frac{6pGM_0}{c^2 a(1-e^2)} \quad (45)$$

or, in other words, the perihelion rotates in the retrograde sense.

**Discussion**

We just derived the perihelion precession by postulating a solar cogravitational field. According to general relativity, the sun should have indeed a cogravitational field, but due to its spin, not to relative orbital motion. Yet, the classical general relativity derivation of the perihelion precession does not attribute the effect to a cogravitational-type effect but instead to space curvature because of the space-space $(g_{ij})$ components of the metric tensor used in the Schwartzschild metric which is a static metric. In general relativity a static metric is one which has no non-vanishing off-diagonal components in the metric. Thus, in general relativity, a static metric should not be able to generate a cogravitational field. In other words it is completely time invariant and is not moving in any way. So it is really surprising that the inherently dynamic results we present above reproduce a purely static general relativity result.

Indeed, in general relativity, the lowest level of complication of the metric tensor for which a cogravitational vector potential can appear, would be for a stationary metric, which has non-vanishing space-time $(g_{i0})$ components. In other words, the general relativity cogravitational vector potential does not vanish for a stationary metric. A stationary metric, for example, corresponds to constant, uniform rotation. The Kerr solution of general relativity is such a metric. The sun is a good example of a body which generates an approximately stationary metric for its external gravitational field.

In gravitocogravitism the cogravitational field gets important for speeds non negligible compared to the speed of light. This is the diametric opposite of the regime of validity for linearized GR! The linearized GR theory is valid in the low speed and small mass regime. Moreover the results one obtains with GC are impossible to recover in the low velocity small masses case of linearized GR. The interpretation of these facts fall out of the scope of the present paper.

## *Conclusion*

From the rational above we can conclude that we have two possibilities:

- We may observe an advance of the perihelion when the planet is rotating in the prograde sense, and a retardation when the planet is rotating in the retrograde sense.
- Or we observe a retardation of the perihelion when the planet is rotating in the prograde sense, and an advance when the planet is rotating in the retrograde sense.

In our universe we observe the first case. However no measurements have been done with planets rotating in the retrograde sense. This would be important to do, in order to check whether we observe a retardation when the sense of rotation is retrograde. It is important to note that general relativity does not predict this effect.

Even thought there are no retrograde planets, there are known to be retrograde moons about other planets in our solar system. These small bodies might be very difficult to test, but in principle it might be possible.

## *Acknowledgement*